\documentclass[review]{elsarticle}

\usepackage[left=3cm, right=3cm, top=2.5cm, bottom=2.5cm]{geometry}

\usepackage{lineno,hyperref}
\modulolinenumbers[1]
\usepackage{caption}
\usepackage{subcaption}
\usepackage{xcolor}
\usepackage{graphicx}
\usepackage{amssymb}
\usepackage{relsize}
\usepackage{float}

\usepackage{subcaption} 

\begin{document}

\begin{frontmatter}

\title{Prediction of local elasto-plastic stress and strain fields in a two-phase composite microstructure using a deep convolutional neural network} 

\affiliation[inst1]{organization={Department of Civil and Systems Engineering, Johns Hopkins University},
            addressline={\\ 3400 N. Charles Street}, 
            city={Baltimore},
            postcode={21218}, 
            state={Maryland},
            country={USA}}
\author[inst1]{Indrashish Saha}
\author[inst1]{Ashwini Gupta}
\author[inst1]{Lori Graham-Brady\corref{mycorrespondingauthor}}


\cortext[mycorrespondingauthor]{Corresponding author.\\E-mail Address:  \href{mailto:lori@jhu.edu}{lori@jhu.edu} (L. Graham-Brady).}

\begin{abstract}

Design and analysis of inelastic materials requires prediction of physical responses that evolve under loading. Numerical simulation of such behavior using finite element (FE) approaches can call for significant time and computational effort. To address this challenge, this paper demonstrates a deep learning (DL) framework that is capable of predicting micro-scale elasto-plastic strains and stresses in a two-phase medium, at a much greater speed than traditional FE simulations. The proposed framework uses a deep convolutional neural network (CNN), specifically a U-Net architecture with 3D operations, to map the composite microstructure to the corresponding stress and strain fields under a predetermined load path. In particular, the model is applied to a two-phase fiber reinforced plastic (FRP) composite microstructure subjected to a given loading-unloading path, predicting the corresponding stress and strain fields at discrete intermediate load steps. A novel two-step training approach provides more accurate predictions of stress, by first training the model to predict strain fields and then using those strain fields as input to the model that predicts the stress fields. This efficient data-driven approach enables accurate prediction of physical fields in inelastic materials, based solely on microstructure images and loading information.


\end{abstract}


\end{frontmatter}






\section{Introduction}

Mechanical failure of composite materials is associated with localization - high local stresses developing in the material~\cite{llorca2011multiscale, chen2021advances, nepal2023hierarchically}. Therefore, prediction of local physical responses is crucial in computational design and analysis of composite materials. In inelastic materials, this requires a non-linear analysis to determine the critical stresses developing in the composite microstructure under loading. Conventionally, this non-linear analysis is performed by  using numerical approaches such as finite element (FE) methods~\cite{reddy2019introduction, belytschko2014nonlinear, simomarsden1998}. However, these numerical approaches often require significant time and computational effort because they call for many iterations to solve the  partial differential equations that describe computational plasticity. Additionally, the behavior of fiber reinforced plastic (FRP) composite materials depends on the random arrangements of fibers in their microstructure~\cite{torquato2002random, matsuda2003effects, kim2010elastoplastic}, so a new analysis is needed for each specific fiber arrangement. 

Over the past decades, data-driven surrogate models have emerged as practical alternatives to physics-based models, to mitigate these computational expenses~\cite{cristianini2000introduction, xu2005stochastic, paulson2017reduced, teferra2018random, bhaduri2021probabilistic}. Among the surrogate models, machine learning (ML) based models have become particularly popular ~\cite{bock2019review, stoll2021machine, kovachki2022multiscale}. The success of many ML approaches in modeling structure-property relations relies on efficient microstructure representation, or in other words the extraction of information-rich low-dimensional features from high dimensional microstructure images~\cite{jiao2007modeling, brough2017materials, pathan2019predictions, li2022machine}. Recent works have shown that convolutional neural networks (CNNs) are an ideal choice for autonomous microstructure characterization as they utilize convolutional layers to automatically learn hierarchical features directly from images~\cite{li2018transfer, bhaduri2021efficient, guo2021artificial}. Significant efforts have been devoted to utilizing CNNs for predicting effective mechanical properties of materials~\cite{yang2019using, chen2022data}. For example, Sengodan~\cite{sengodan2021prediction} has combined dimension reduction methods and CNNs to predict the homogenized stress-strain curve of inelastic two-phase microstructures. Kim et al.~\cite{kim2021prediction} have directly used the high-dimensional microstructure as the input to a CNN model that predicts the homogenized stress-strain curve. The primary goal of these homogenization-based CNN models is to establish a mapping between the composite microstructure and its macroscopic properties. However, these models do not tackle the issue of localization - high local stresses that develop within the composite microstructure.

Recent advancements have shown the effective utilization of CNNs for predicting local stress fields in elastic composites~\cite{yang2019establishing, yang2021end, rashid2022learning, park2022generalizable}. Bhaduri et al.~\cite{BHADURI} have used a U-Net CNN to map the composite microstructure with varying number of fibers to the elastic von Mises stress field. Gupta et al.~\cite{gupta2023accelerated} have used a U-Net CNN in a fast multiscale analysis approach that performs both homogenization and localization. This ML-driven multiscale approach uses a single CNN to predict the stress tensor fields for elastic composites under a general state of loading. The challenge in extending this ML-driven multiscale analysis framework to address inelastic behavior is the evolution of such systems under accumulating plastic strains. Sepasdar et al.~\cite{sepasdar2022data} have used a sequence of CNNs to predict post-failure von Mises stress field and the failure patterns in inelastic composites. The majority of these past works employing CNNs considers only an elastic framework or focuses solely on prediction of stress/strain fields at the end of the loading path. These image-to-image mapping approaches face limitations in predicting intermediate local stress fields under plastic deformation.

In the current work, we present a deep-learning (DL) framework that is capable of predicting the evolution of micro-scale elasto-plastic strains and stresses in a two-phase medium. The proposed DL-framework uses a convolutional neural network (CNN) to map the composite microstructure to the corresponding stress and strain tensor fields under a predetermined load path. This DL-framework uses 3D U-Nets to model the elasto-plastic responses as a sequence prediction problem. This approach is demonstrated through application to a two-phase FRP composite microstructure subjected to a predetermined loading-unloading path. A novel two-step training approach provides more accurate predictions, by first training the model to predict strain fields and then using those strain fields as input to the model that predicts the stress fields. This efficient data-driven approach enables accurate prediction of physical fields in inelastic materials at a much faster speed than conventional FE simulations.

\indent This paper is structured as follows: Section \ref{sec:methodology} describes the proposed DL-framework for predicting micro-scale elasto-plastic strains and stresses in a composite material. Section \ref{sec:results} describes the results obtained using this method and Section \ref{sec:conclusions} presents the conclusions made from this study.


\section{Methodology}\label{sec:methodology}
\subsection{Inelastic composite problem statement}
\begin{figure}
\centering
         \includegraphics[width=0.6\textwidth]{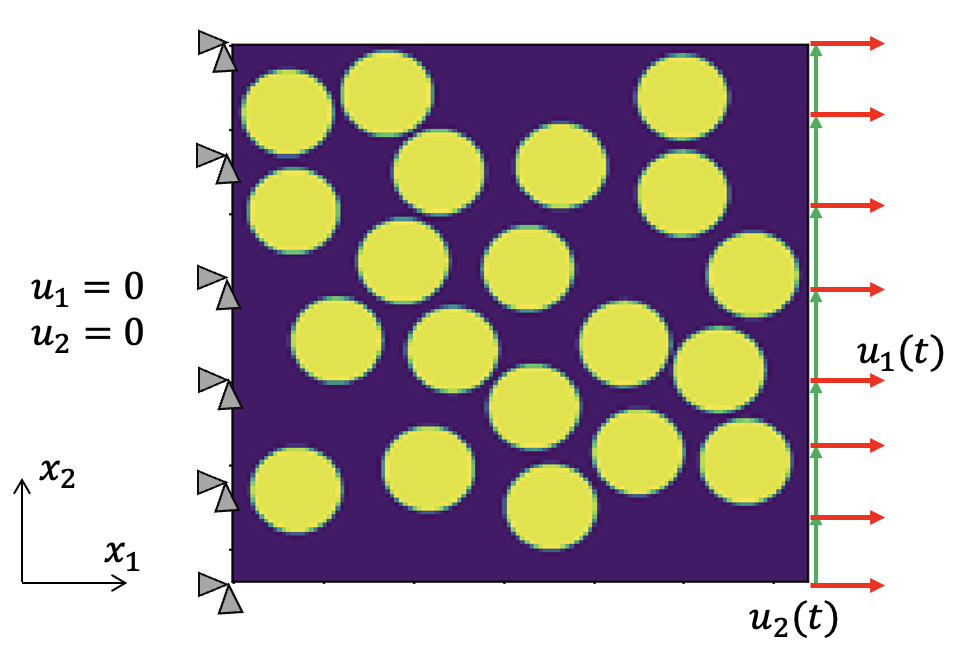}
         \caption{Boundary conditions assumed for the microstructure}
         \label{fig:Boundary condition}
\end{figure}

\begin{figure}
\centering
         \includegraphics[width=0.45\textwidth]{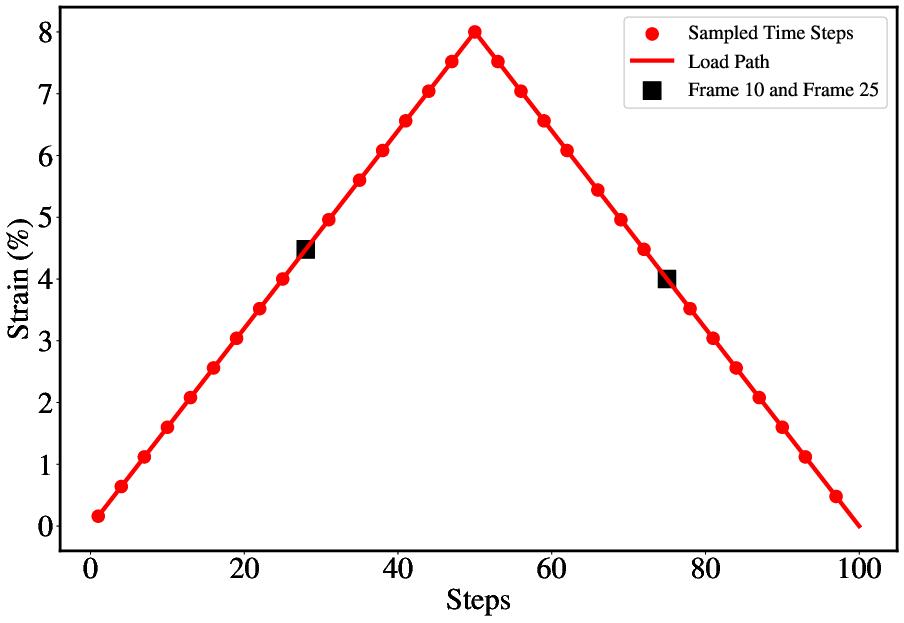}
         \caption{Path followed for loading and unloading the microstructure for $u_{1}(t)$ and $u_{2}(t)$}
         \label{fig:Actual load path}
\hfill 
\end{figure}

This work aims to predict the strain tensor field $\epsilon_{ij}(x,y,t)$ and the stress tensor field $\sigma_{ij}(x,y,t)$ in a 2D two-phase microstructure $M(x,y)$ that is loaded under a combination of uniaxial tensile ($u_1(t)$) and shear ($u_2(t)$) displacement (see Fig.~\ref{fig:Boundary condition}). Specifically, the microstructure consists of randomly distributed stiff circular fibers (fully elastic) in an elastic-perfectly plastic matrix material. The fibers have a fixed diameter, the fiber volume fraction ($V_f$) is kept constant at 40\%, and the fiber/matrix interface is assumed to be perfectly bonded. The material constants for both the matrix (epoxy) and the fiber (glass fiber) are given in Table ~\ref{material}~\cite{Material_prop}. A single cycle of loading and unloading is applied to the microstructure, as illustrated in Fig.~\ref{fig:Actual load path}, to capture the inelastic behaviour of the material. A maximum of $8\%$ strain is applied in both shear and tension, which is sufficient to yield the matrix (epoxy) ~\cite{Material_prop}. 
\begin{table}[h!]
\centering
\caption{Material constants assumed for the components of the two-phase material}
\begin{tabular}{ c | c | c |c}
 Component & $E$ (MPa) & $\nu$ & $\sigma_{y}$ (MPa)  \\ 
 \hline
Epoxy & 3200.0 & 0.3 & 210.0 \\ 
Glass Fiber & 87000.0 & 0.2 &  ---
\end{tabular}
\label{material}

\end{table}

\indent FE simulations via ABAQUS provide the stress $\sigma_{ij}^{(k)}(x,y)$ and strain fields $\epsilon_{ij}^{(k)}(x,y)$ in the material under the given loading conditions at specific load steps $t_k$ along the load path, illustrated in Fig.~\ref{fig:Actual load path}. 
The same load path is applied for both shear and tensile strains simultaneously, so that the applied shear and the tensile strains are equal at every time step. In this analysis, the number of specific load steps under consideration is $K=32$, indicated by the red circles on the load path. The black squares indicate specific instants on the load path for which detailed visualizations will be provided in this paper.  
Note that the FE analysis used to generate the training data considers more finely spaced load steps, but this subset of 32 frames is extracted for subsequent analyses via the ML model. As shown in Fig.\ref{fig:Boundary condition}, the left edge of the microstructure is assumed to be pin-supported, while a uniaxial displacement ($u_1(t)$) and a shear displacement ($ u_2(t)$) are applied to the right edge. 
The top and bottom boundaries are assumed to be traction-free. The matrix material is assumed to be elastic-perfectly plastic with a von-Mises yield criterion and an assumption of small strain $J_2$ plasticity ~\cite{simomarsden1998}. The FE model uses an irregular mesh, with $86000$ elements on average for every microstructure
, finely resolving the distribution of stresses and strains around and between fiber-matrix interfaces. After the analysis is complete, the input microstructure $M(x,y)$ is discretized into an $N\times N$ square pixel image $M^{(mn)}$, and output stress and strain fields at each load step $k$, $\sigma_{ij}^{(k)}(x,y)$ and $\epsilon_{ij}^{(k)}(x,y)$, are similarly discretized into $N\times N$ square pixel images. The stress and strain fields therefore become 3D voxelized images $\sigma_{ij}^{(mnk)}$ and $\epsilon_{ij}^{(mnk)}$, where $(mnk)$ indicates the value at point $(x_m,y_n)$ during load frame $t_k$. For the current analysis, the image size $N$ is taken to be 128, which is a standard size for many images. The discretized $128\times 128$ microstructure and three-dimensional $128\times 128 \times 32$ images of stresses and strains serve as the basis for subsequent training of the 3D U-Net model, as described in the next subsection. 
\subsection{3D U-Net architecture}
\begin{figure}[h!]
  \centering
         \includegraphics[width=\textwidth]{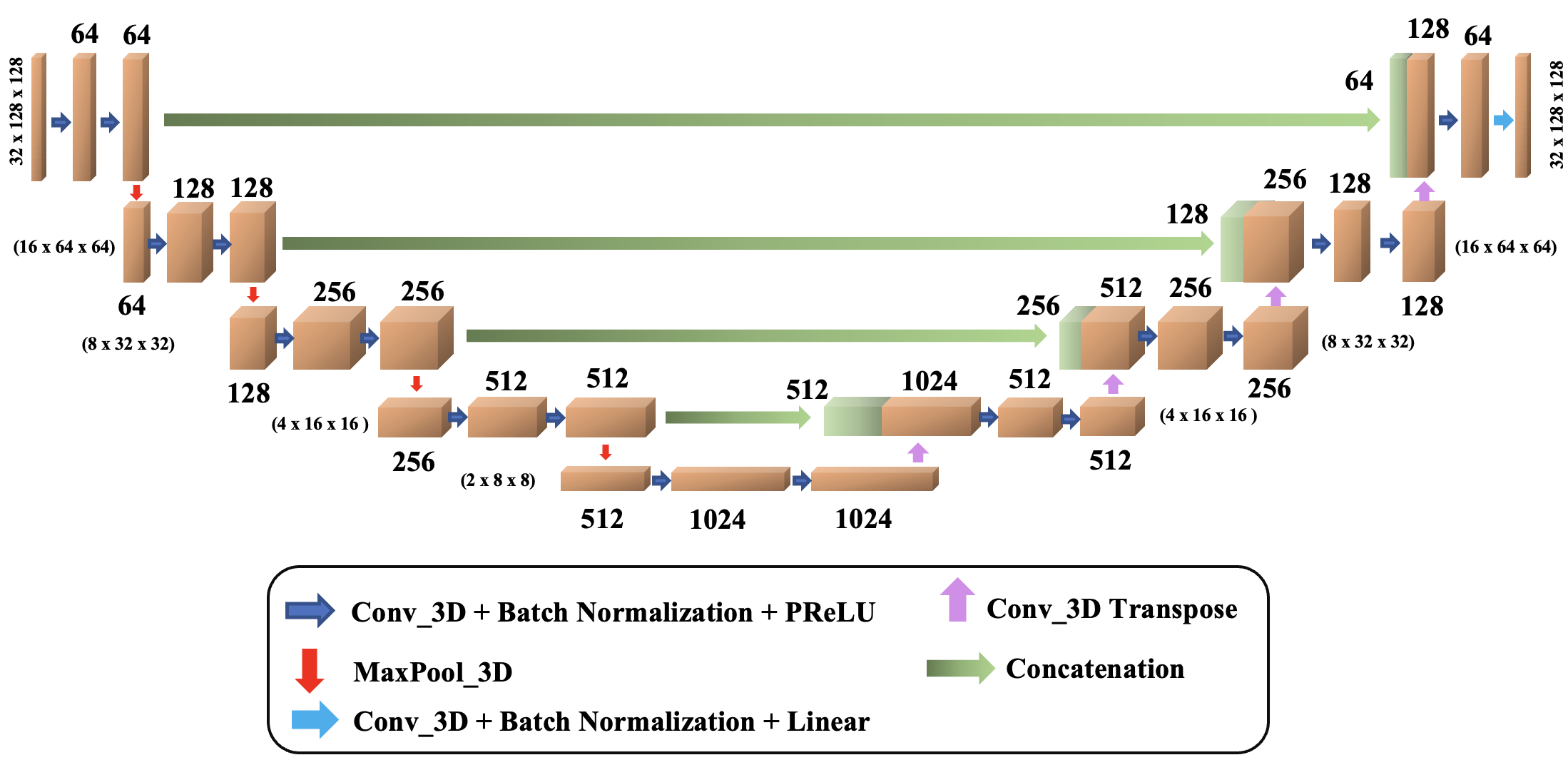}
         \caption{3D U-Net architecture for 32 frames of image}
         \label{fig:UNet}
\end{figure}

\begin{figure}[h!]
\subcaptionbox{PReLU\label{fig: prelu}}{\includegraphics[width=0.5\textwidth]{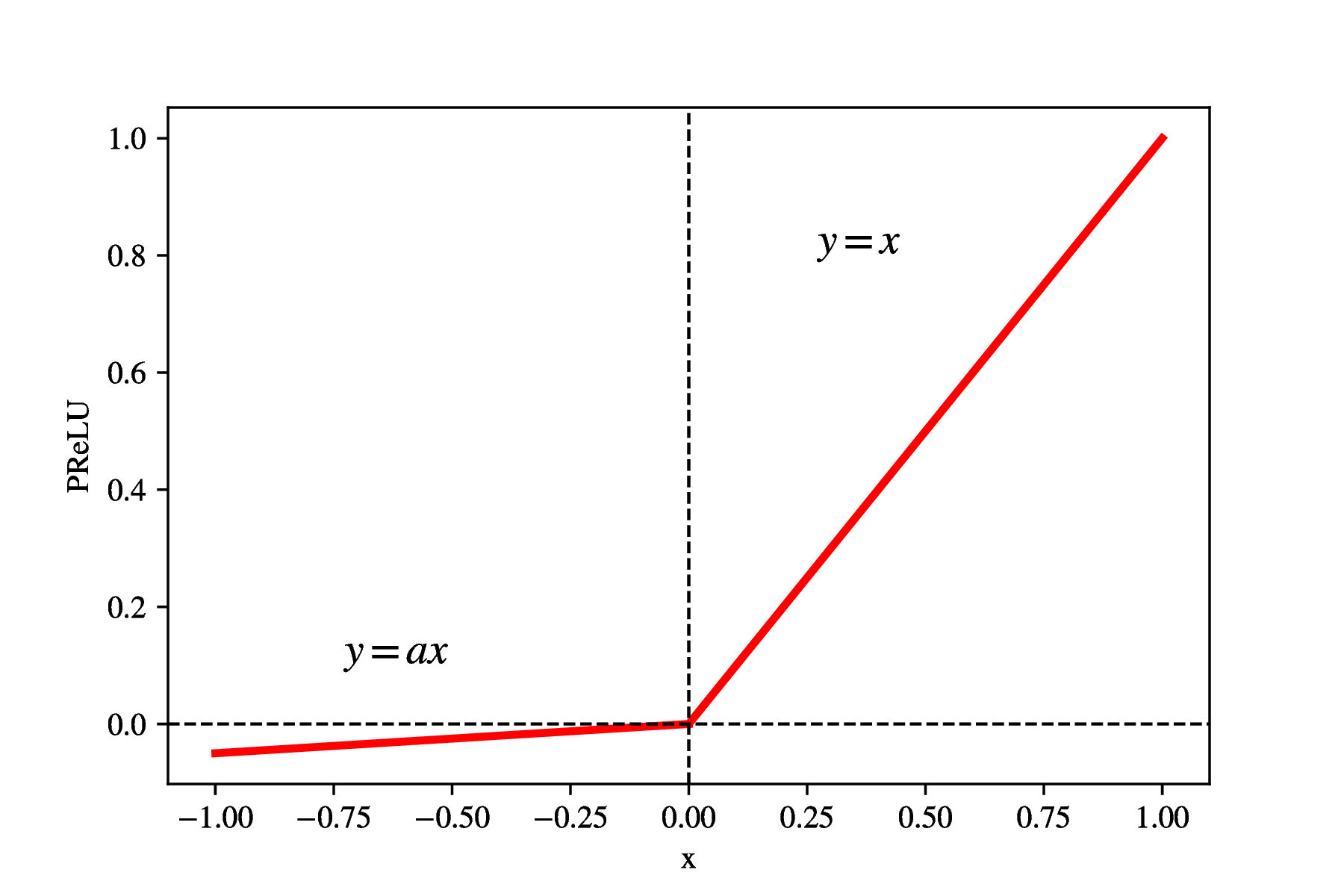}}
\subcaptionbox{Linear\label{fig: linear}}{\includegraphics[width=0.5\textwidth]{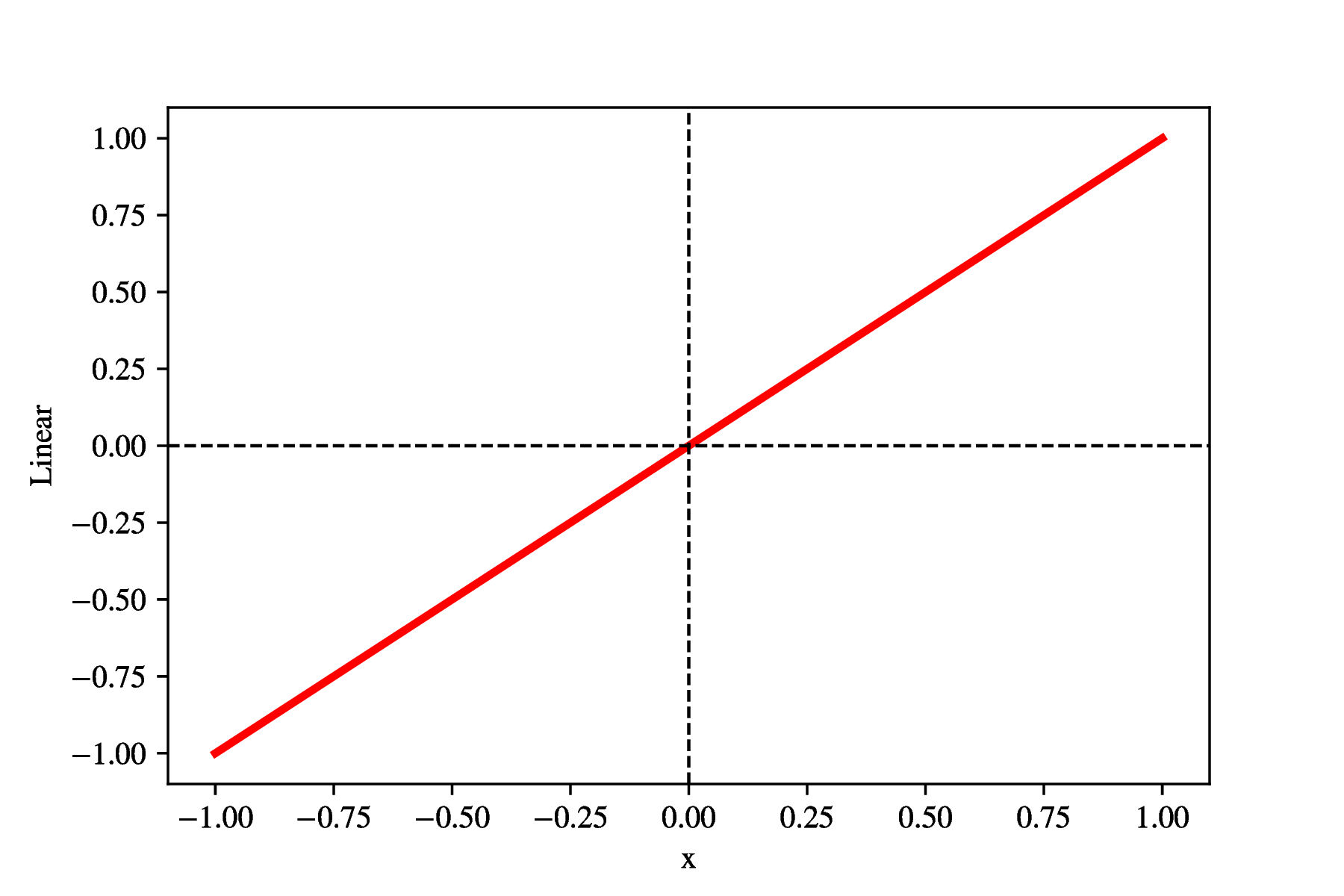}}%
\caption{Activation functions used in the U-Net architecture}
\label{fig:activation}
\end{figure}

\indent The U-Net architecture containing 2D convolutions is an established method for segmentation of images in computer vision ~\cite{ronneberger2015u}. It has an encoder-decoder type architecture with additional skip connections between the encoder blocks and decoder blocks. The skip connections provide additional spatial information to the decoding layers which results in a better quality reconstruction of the spatial features. In the domain of computational mechanics, the 2D U-Net architecture has proven to be effective in the prediction of elastic stress maps ($\sigma_{ij}(x,y)$) from 2D microstructure images~\cite{BHADURI} by solving a regression problem. 
\\ 
\indent Plasticity is a history-dependent material behavior, meaning that the current state of the material is dependent on its previous states. For example,
 \begin{figure}
 \centering
          \includegraphics[width=0.5\textwidth]{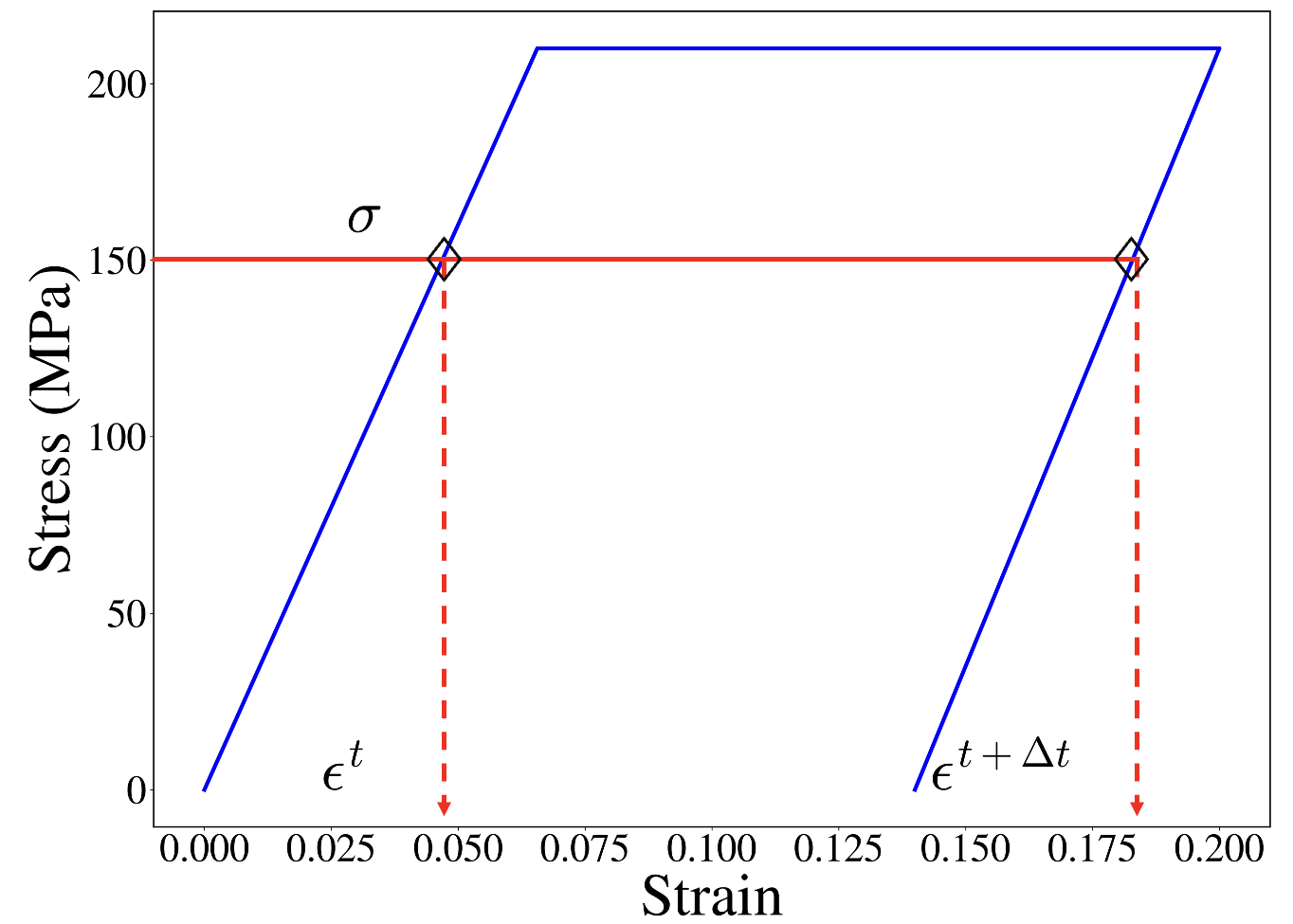}
          \caption{Typical stress-strain response of a material undergoing loading and unloading in the plastic regime}
          \label{fig:plasticity}
 \end{figure}
in Fig.~\ref{fig:plasticity}, 
two strain states ($\epsilon^{t}$ and $\epsilon^{t+\Delta t}$) are possible for a given stress state $\sigma$. The superscripts $t$ and $t+\Delta t$ refer to the sequence of observation of the strain states. Therefore, in order to predict a particular $\sigma_{ij}(x,y)$ or $\epsilon_{ij}(x,y)$ in the plastic regime, the stress and strain histories are required. Thus, the stresses ($\sigma_{ij}(x,y,t)$) and the strains ($\epsilon_{ij}(x,y,t)$) become a function of time. The 2D U-Net architecture is only capable of working with the current state of stress ($\sigma_{ij}(x,y)$). In order to improve upon this, a 3D U-Net is used in this work, due to its ability to represent sequential information in the third dimension.\\
\indent 
Similar to the 2D U-Net architecture, the 3D U-Net is comprised of a down-sampling and an up-sampling path, as illustrated in Fig.~\ref{fig:UNet}. 
Each level on the down-sampling path contains two $3\times 3\times 3$ convolutions and two non-linear activation units known as Parametric-rectified Linear Unit (PReLU) ~\cite{he2015delving}, followed by a $2\times 2\times 2$ max pooling down to the next level. Each level on the up-sampling path consists of a transpose convolution of $2\times 2\times 2$, followed by two $3\times 3\times 3$
convolutions, each followed by PReLU activation. Skip connections between down-sampling and up-sampling levels of similar dimensions transfer high-resolution spatial information to the up-sampling path. In addition to these operations, batch normalization (BN) is added before every PReLU, to accelerate the training process, by helping achieve convergence at a faster rate~\cite{BN_basic}. Since the problem is treated as a regression problem, with both positive and negative values potentially associated with each pixel, a PReLU is applied in the intermediate layers and a linear activation function in the final layer, illustrated in Fig.~\ref{fig:activation}. The PReLU activation function is an improvement from the traditional ReLU function which converts all negative values to zero. This feature in the context of a neural network implies that none of the neurons are turned off, because PReLU allows for small negative values. The slope $a$ corresponding to $x<0$ in PReLU, as shown in Fig.~\ref{fig: prelu}, is a parameter that is learned during the training, which essentially fits a different value for each neuron. Thus, PReLU allows improved adaptability of the network with the data, which in turn results in better convergence.

The loss function used in training the U-Net for any given 3D field $z^{(mnk)}$ is chosen as the mean absolute error (MAE) between the predicted and true images,
\begin{equation}
 MAE=\frac{\sum_{m=1}^{N}{\sum}_{n=1}^{N}\sum_{k=1}^K|z_{true}^{(mnk)}-z_{pred}^{(mnk)}|}{N^2K}
    \label{eq:1}
\end{equation}
   where $K$ is the number of load frames for which the $N\times N$ images are available,  and $z_{pred}^{(mnk)}$ and $z_{true}^{(mnk)}$ are the pixel values at location $(x_m,y_n)$ and load frame $t_k$ of the predicted image and true image, respectively. For the purposes of this study, the learning rate of the gradient descent algorithm was selected to be 0.001 and the exponential decay rate for the first moment estimates of the Adam optimizer was set at 0.96~\cite{3D_Unet_basic}. While these hyperparameters can be modified, this particular choice of the learning rate, the optimizer and its parameters are fairly standard for U-Net architectures.

\subsubsection{Arrangement of the U-Net framework}
\begin{figure}[h!]
  \centering
         \includegraphics[width=\textwidth]{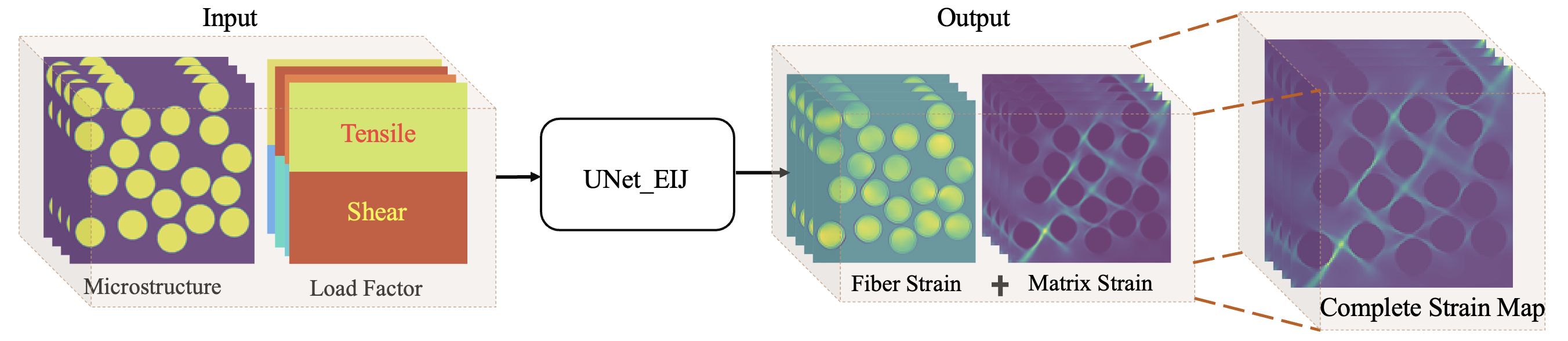}
         \caption{Schematic representation of UNet\_EIJ. The input channel(s) contain the microstructure and optionally a color map that indicates the applied displacements at each load step (both axial and shear). The output channel could contain the complete strain map $\epsilon_{ij}^{(mnk)}$ directly, or it could include separate channels for fiber strains $\epsilon_{ij}^{(F,mnk)}$ and matrix strains  $\epsilon_{ij}^{(M,mnk)}$ that are subsequently brought together to predict the overall strain $\epsilon_{ij}^{(mnk)}$. 
         }
         \label{fig:EIJchart}
\end{figure}

\indent As illustrated in Fig.~\ref{fig:EIJchart}, three independent U-Net models provide direct mappings from the given microstructure $M^{(mn)}$ to each component $ij$ of the strain field $\epsilon_{ij}^{(mnk)}$. `UNet\_EIJ' (I,J $\epsilon \{1,2\}$) is adopted as the nomenclature to represent the U-Net model that predicts the (I,J) components of strain. The input channels may include only the microstructure, but the results in Section \ref{sec:results} will show that accuracy is improved significantly by adding another input channel with an image that explicitly indicates the magnitude of applied displacement at each load step. 
There are two options for the strain output channels: 1) direct output of the overall strain map $\epsilon_{ij}^{(mnk)}$; or 2) direct output of fiber strains $\epsilon_{ij}^{(F,mnk)}$ and matrix strains $\epsilon_{ij}^{(M,mnk)}$ separately, then subsequently joining them together into an overall strain map $\epsilon_{ij}^{(mnk)}$. In the second approach, 
the strains are separated by taking advantage of the microstructures $M^{(mn)}$, binary images in which $1$ represents fiber region and $0$ represents the matrix region. Therefore, for any load step $k$, a set of scaled strains associated with the fibers is found as: 
\begin{equation}
    \epsilon_{ij}^{(F,mn)} = \lambda_F M^{(mn)}\circ\epsilon_{ij}^{(mn)}
    \label{fibre strain}
\end{equation}

\noindent
and similarly a set of scaled strains associated with the matrix is found as:

\begin{equation}
    \epsilon_{ij}^{(M,mn)} = \lambda_M (\mathbb{J}-M^{(mn)})\circ\epsilon_{ij}^{(mn)}
    \label{matrix strain}
\end{equation}
where $\mathbb{J}$ represents a matrix whose elements are ones and '$\circ$' represents element-by-element multiplication. $\lambda_F$ and $\lambda_M$ are scaling factors for the fibre and matrix strains respectively. Finally, the predicted strain maps can be reconstructed as

\begin{equation}
    \epsilon_{ij}^{(mnk)} = \frac{\epsilon_{ij}^{(F,mnk)}}{\lambda_F}+\frac{\epsilon_{ij}^{(M,mnk)}}{\lambda_M}
    \label{total strain}
\end{equation}

\noindent After testing some trial values for $\lambda_F$ and $\lambda_M$, it was determined that $10^3$ and $10^4$, respectively, led to accurate predictions. 
 However, no formal studies were conducted to evaluate the effect of the constants on the error. The results from direct output of the total strain map and from direct output of separate fiber and matrix strains will be evaluated in the results in Section \ref{sec:results}. 

\begin{figure}[h!]
  \centering
       \includegraphics[width=\textwidth]{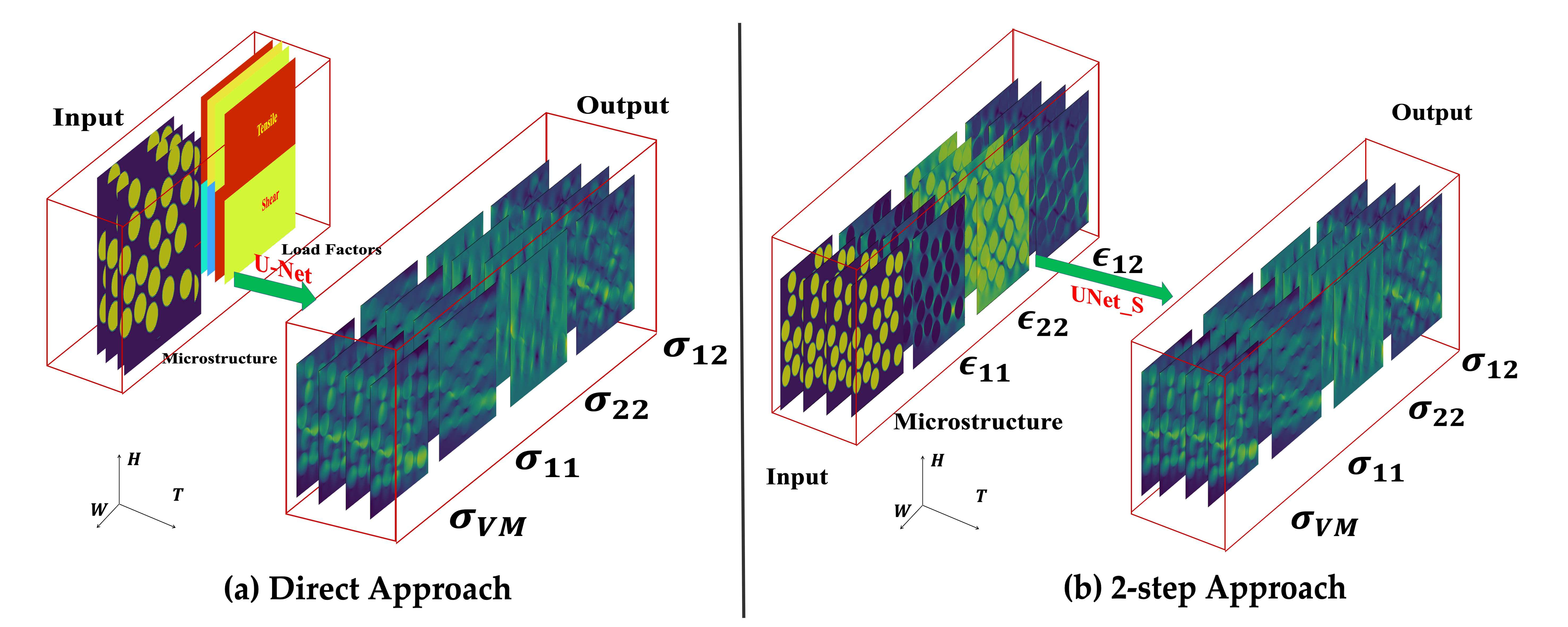}
         \caption{A typical 3D data arrangement for the UNet\_S architecture; $H =$ Height of the image ; $W =$ Width of the image; $T =$ Temporal information stored as the third dimension in the data unit. Two approaches to calculate stress are considered: (a) direct prediction of stresses based on microstructure and load factors, and (b) prediction of stresses based on microstructure and the U-Net-predicted strain fields.  
         }
         \label{fig:data unit}
          \hfill
\end{figure}

Four stress fields ($\sigma_{ij}^{(mnk)}$ and von-Mises stress $\sigma_{VM}^{(mnk)}$) are predicted using a single U-Net (`UNet\_S') with multi-channel output. 
As illustrated in Figure \ref{fig:data unit}, there are two options for training a U-Net to predict the stress fields: 1) training based solely on the microstructure $M^{(mn)}$ as input (direct approach); and 2) training based on the microstructure $M^{(mn)}$ and the strain field predictions from the strain U-Nets $\epsilon_{ij}^{(mnk)}$ as input (2-step approach). These options will be evaluated further in the results in Section \ref{sec:results}. 


\subsubsection{Data Set}
Each data unit has a third dimension, as illustrated in Fig.~\ref{fig:data unit}, corresponding to the
thirty-two points on the load path indicated in Fig.~{\ref{fig:Actual load path}}. In the current work, a total of 420 microstructures were used for training the U-Net models, while 44 additional microstructures were reserved for testing purposes. The testing dataset was specifically chosen to contain microstructures that were not seen during the training phase. This separation allows for evaluating the generalizability of the trained models on unseen data. Additionally, by taking advantage of the symmetry of the problem, the microstructures and the corresponding stress and strain fields were rotated by $180^\circ$ and were added to the dataset. Therefore, the size of the training dataset was doubled and 840 training microstructures were obtained. The size of the dataset was fixed once the predicted mean strain field evolutions had a good agreement with the true values. 
\section{Results and Discussion}\label{sec:results}
\begin{figure}[h!]
  \centering
         \includegraphics[width=\textwidth]{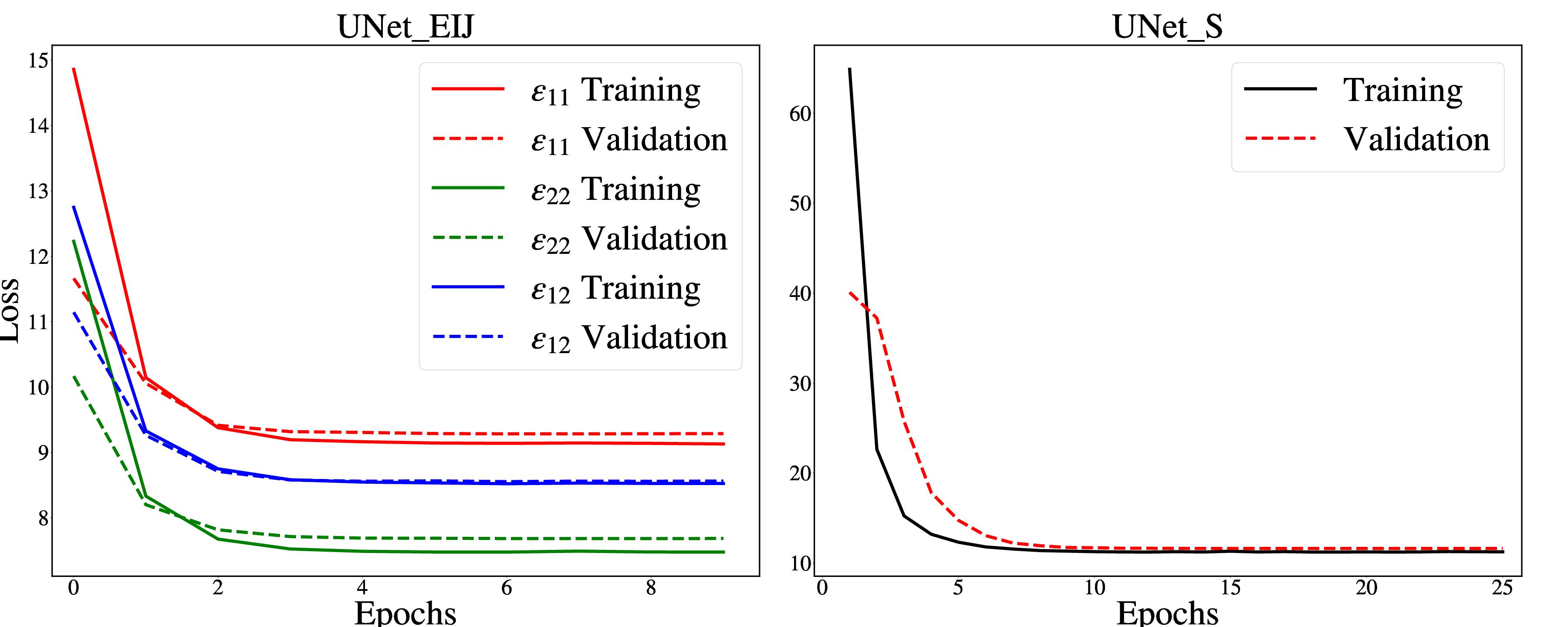}
         \caption{Convergence of the U-Net models}
         \label{fig:loss}
          \hfill
\end{figure}

To ensure that training was performed sufficiently, the UNet\_EIJ networks were trained for 10 epochs and all the models converged in 3 epochs. UNet\_S was trained for 25 epochs and was found to converge in 6 epochs.  These results have been summarised in Fig.~\ref{fig:loss}. Comparing the training and validation results in these figures it is evident that none of the models had overfitting problems.
\subsection{Predicted strain and stress  maps at specific load frames}
\begin{figure}[h!]
  \centering
         \includegraphics[width=\textwidth]{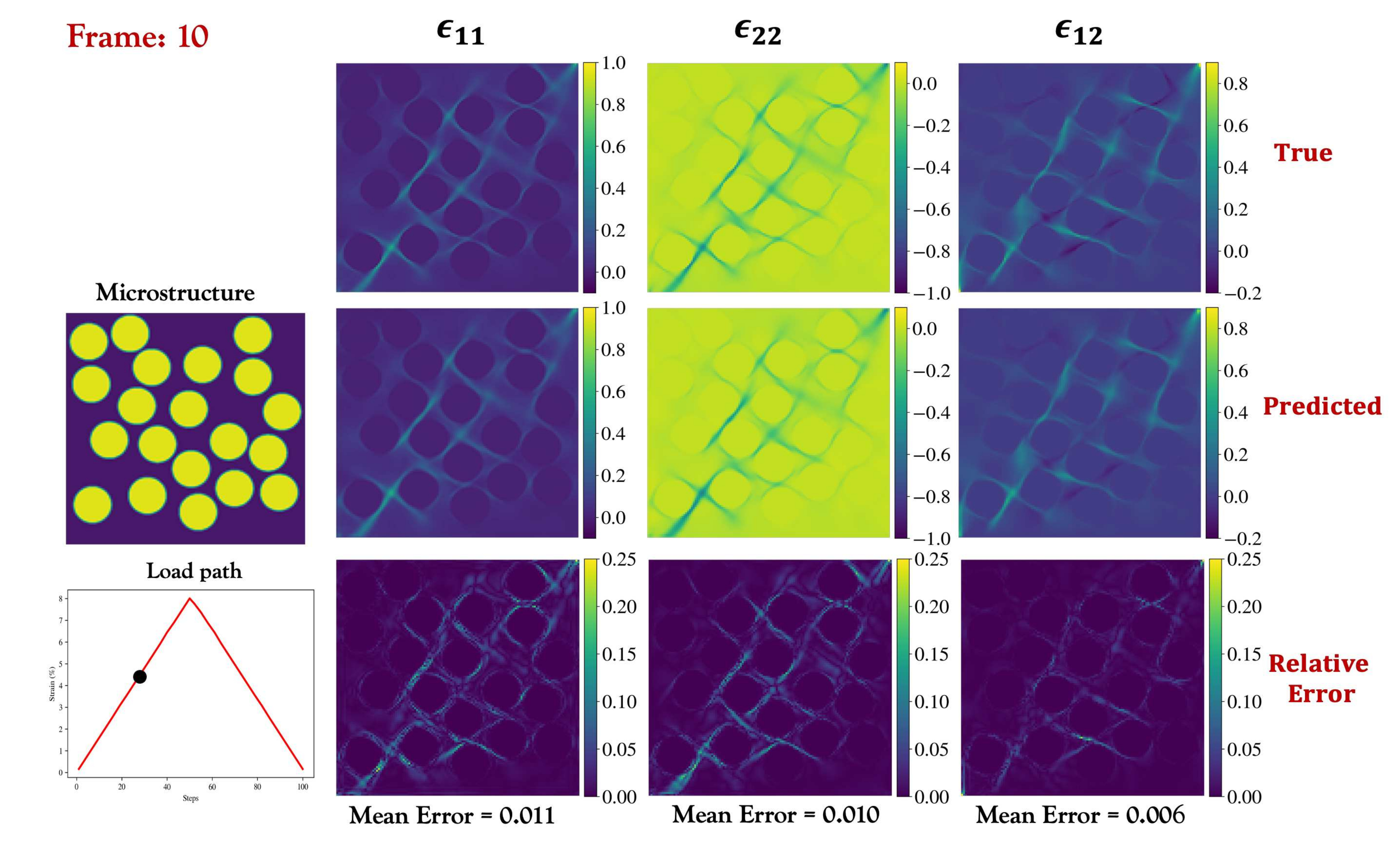}
         \caption{Comparison of the predicted and the true strain maps in the unloading stage (Frame 10)}
         \label{fig:strain10}
          \hfill
\end{figure}

\begin{figure}[h!]
  \centering
         \includegraphics[width=\textwidth]{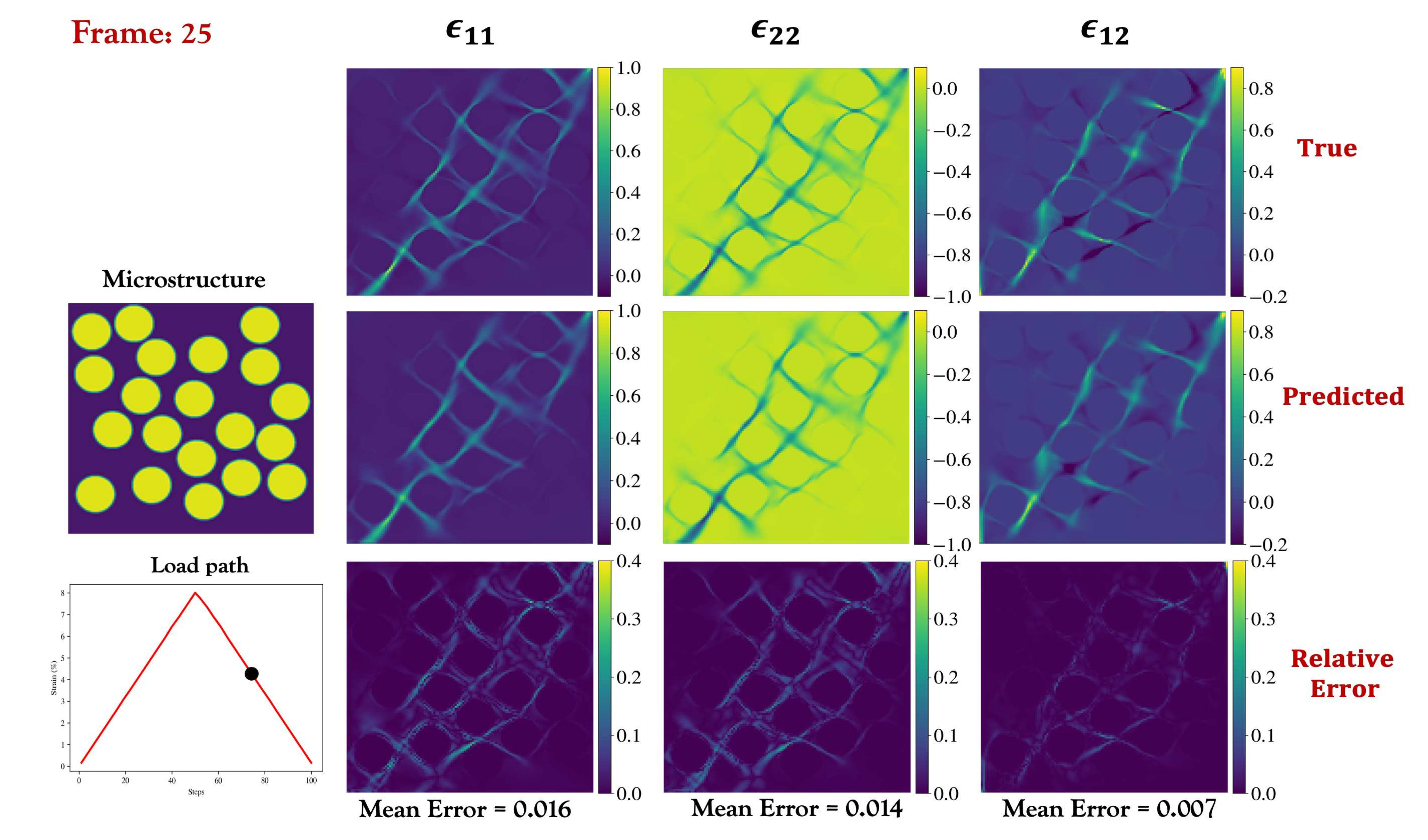}
         \caption{Comparison of the predicted and the true strain maps in the unloading stage (Frame 25)}
         \label{fig:strain25}
          \hfill
\end{figure}

\indent The U-Net models generate stress and strain predictions corresponding to a particular input microstructure, for all 32 load frames. For the purposes of visualization, strain maps are provides for two specific instances: (i) frame 10, on the loading path at $4.5\%$ strain (Fig.~\ref{fig:strain10}), and (ii) frame 25, on the unloading path at $ 4.0\%$ 
strain (Fig.~\ref{fig:strain25}). To evaluate the localised error distribution in the stress/strain maps, these figures show the relative error (RE) for each pixel using, 
\begin{equation}
    RE=\frac{|{z}_{true}^{mn}-z_{pred}^{mn}|}{max\{|{z}_{true}^{mn}|\}}
    \label{Error}
\end{equation}
where ${z}_{true}^{mn}$ and $z_{pred}^{mn}$ are the true and predicted pixel values. Rather than normalizing with respect to each true pixel value in this expression, the normalization is done with respect to the maximum of the true pixel values, to avoid problems at locations where the true pixel value is close to zero. The strain map predictions are very close to the true images, with the RE for frame 10 (loading path) lying in the range of 0-25\% and a maximum average RE of 1.1\% for all the components. 
For frame 25 (unloading path), the RE lies in the range 0-40\%, with an average RE less than 1.6\% for all three strain components.


\begin{figure}[t]
         \includegraphics[width=1\textwidth]{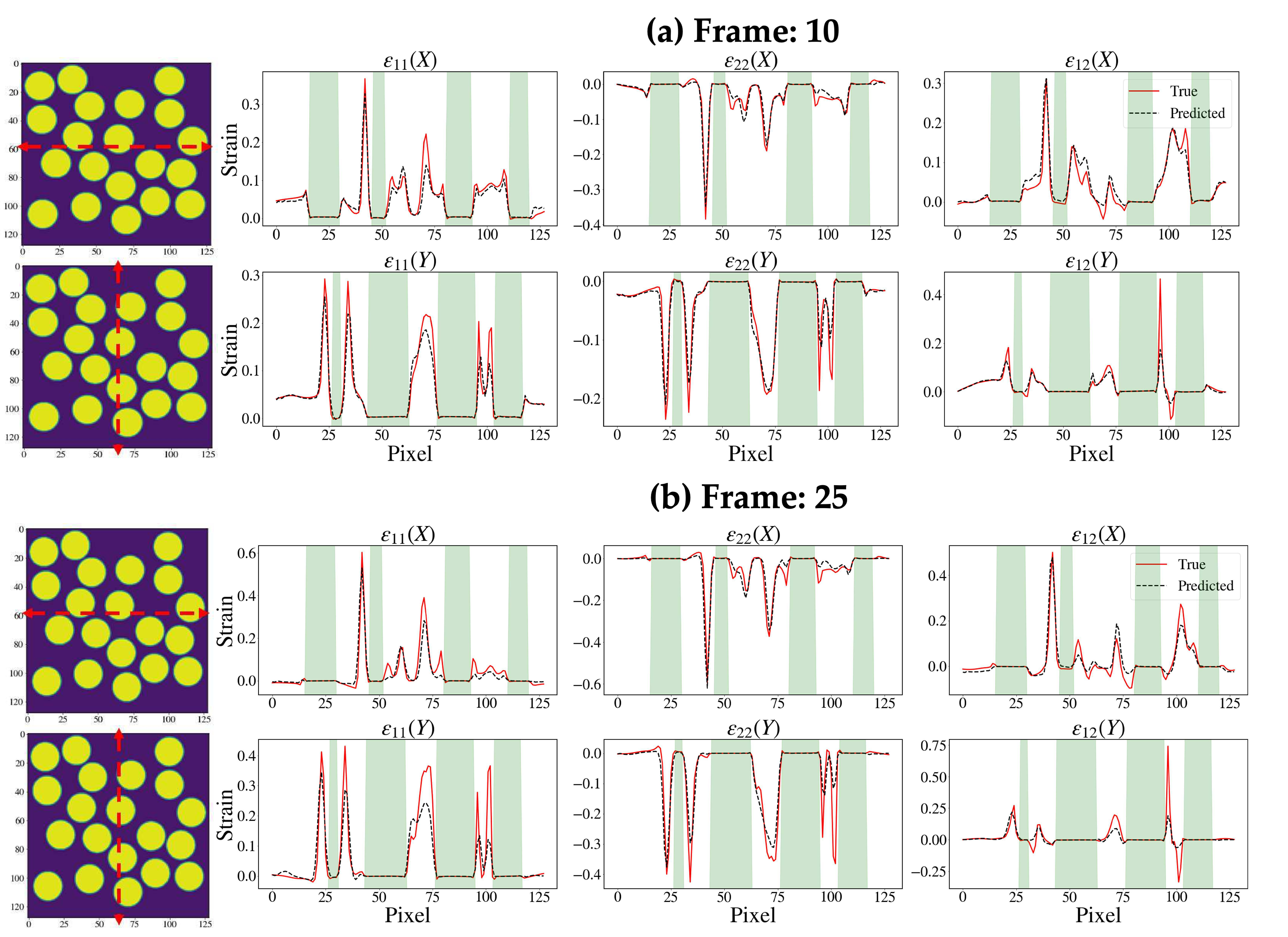}
         \caption{Variation of strains along the midsections. (a) Frame 10 (b) Frame 25 (The green zones signify the regions where the line intersects a fiber) 
         }
         \label{fig:strain_slice}
          \hfill
\end{figure}

To help visualize the local differences between the strain maps, the strains along two lines through the microstructure are provided in Fig.~\ref{fig:strain_slice}a and Fig.~\ref{fig:strain_slice}b, for load frames 10 and 25, respectively. All the strain components show a very good agreement with the predicted strains in both directions. The most significant errors are found in the matrix, near or at fiber-matrix boundaries.  

\begin{figure}[!htb]
  \centering
        \includegraphics[width=1\textwidth]{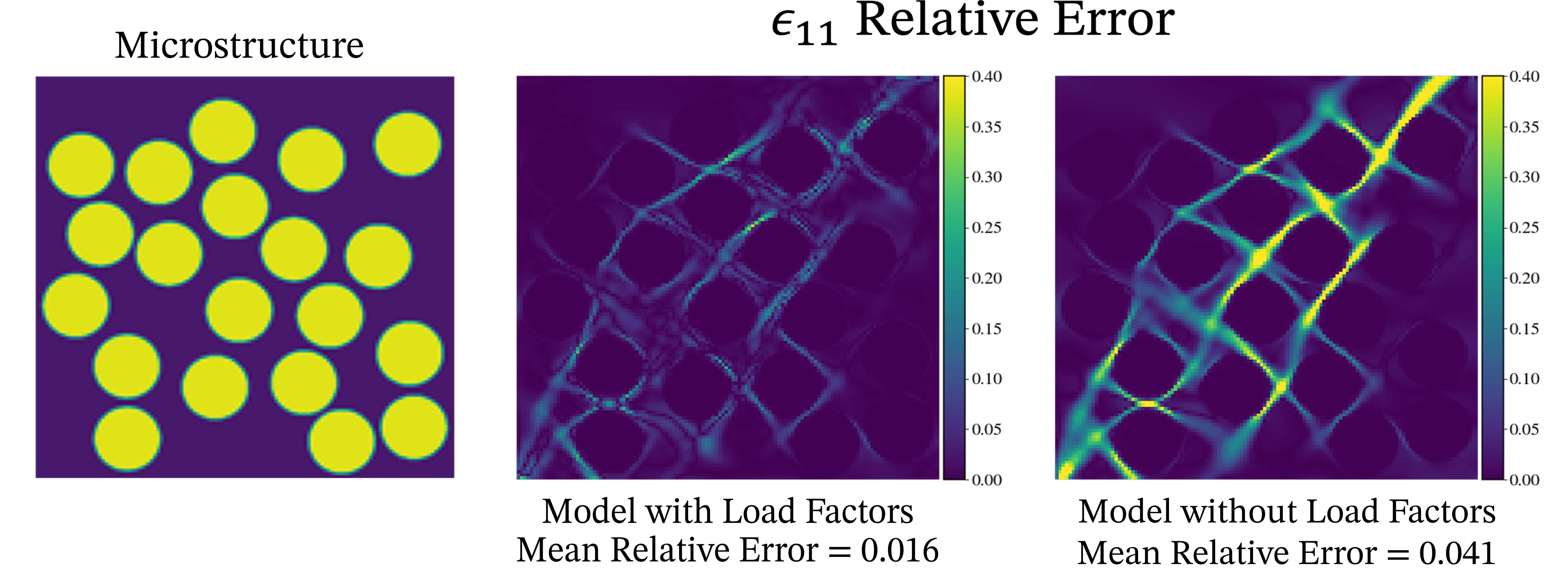}
         \caption{Comparison of the RE in $\epsilon_{11}(x,y)$ at the time step on the unloading path (frame 25), training the strain U-Net model with and without the load factors as explicit input. 
         }
         \label{fig:strain-LF-compare}
          \hfill
\end{figure}

\begin{figure}[h!]
  \centering
         \includegraphics[width=\textwidth]{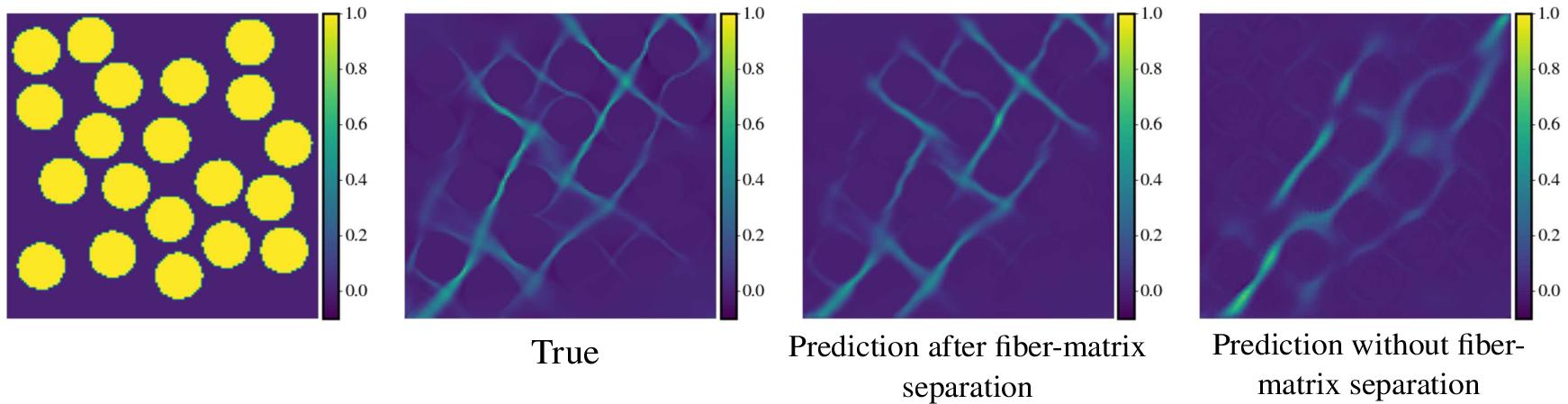}
         \caption{Comparison of the predicted $\epsilon_{11}$ on the unloading path (Image frame : 25) maps with fiber-matrix separation and without fiber-matrix separation.}
         \label{fig:strain_compare}
\end{figure}

\indent As mentioned earlier, input to the U-Net model for strains, UNet\_EIJ, can include images that represent the load factors at each load step. To assess whether or not this is preferable, UNet\_EIJ is trained with and without a 3D image that indicates the applied displacements (axial and shear) at each load step, as illustrated in Fig.~\ref{fig:EIJchart}. Fig.~\ref{fig:strain-LF-compare} shows that the U-Net model with these load factor images as an explicit input leads to an approximately $3\times$ decrease in mean relative prediction error. Therefore, in all other analyses in this current work, these load factors are included in the strain evaluation. 
Also mentioned earlier, the output for each UNet\_EIJ can either include a direct prediction of the strain, or it can include two channels, the fiber strain $\epsilon_{ij}^{F}$ and matrix strain $\epsilon_{ij}^{M}$ that are merged after the analysis. 
As depicted in Figure~\ref{fig:strain_compare}, the predicted $\epsilon_{11}$ maps along the unloading path are displayed for two different approaches: (i) training after separation of fiber and matrix strains (ii) training without the separation of fiber and matrix strain. 
Predictions achieved through the separation of fiber and matrix strains effectively replicate the majority of the high strain regions present in the actual map, while predictions made without the separation of fiber and matrix strains fail to capture high strain regions and exhibit an inaccurate smoothing effect around the fibers. 
Because of the superior performance using separate fiber and matrix strains, all other analyses in this paper are based on this approach. 

\begin{figure}[h!]
  \centering
         \includegraphics[width=1\textwidth]{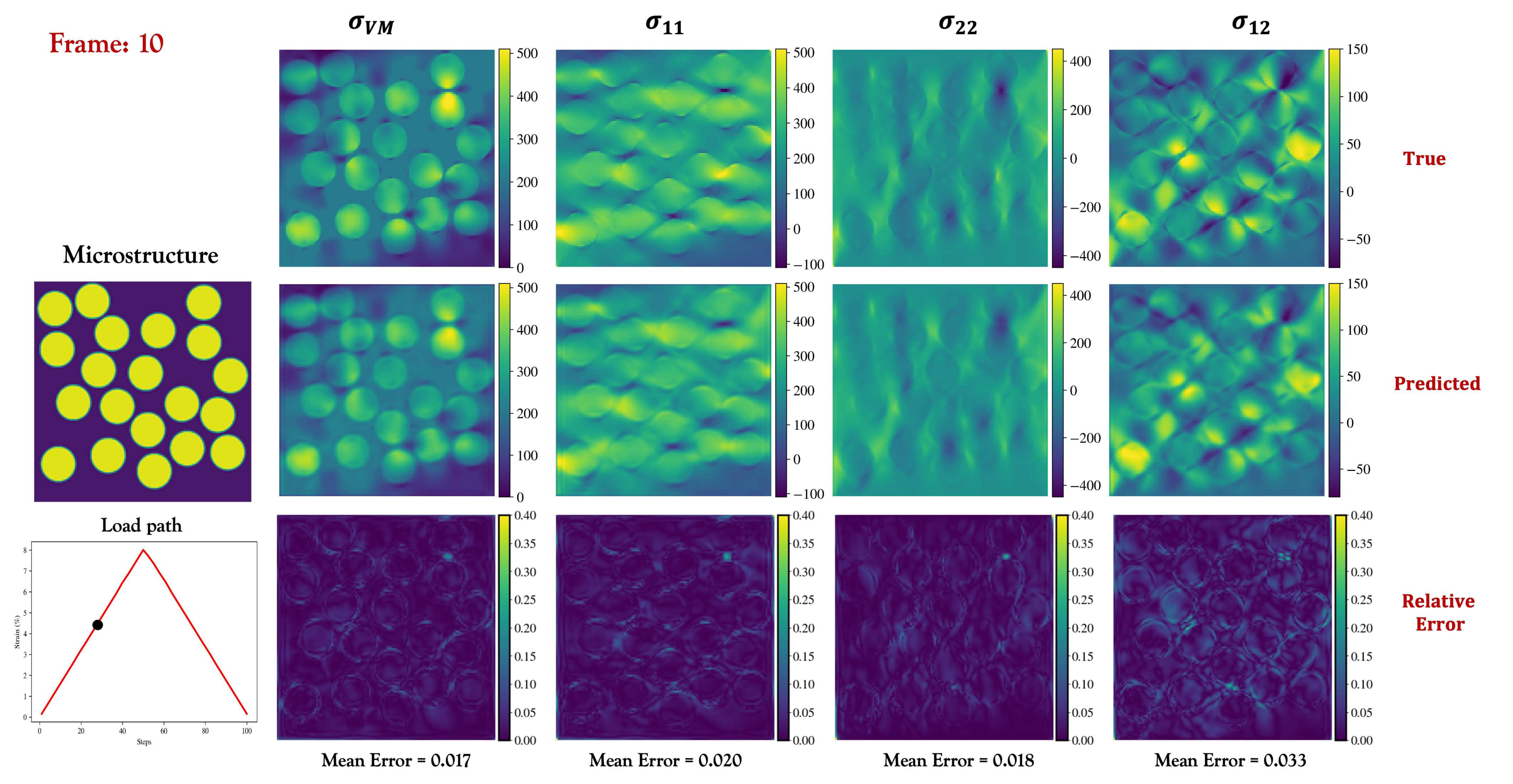}
         \caption{Comparison of the predicted and the true stress maps in the loading stage (Frame: 10)}
         \label{fig:stress10}
          \hfill
\end{figure}

\begin{figure}[h!]
  \centering
         \includegraphics[width=1\textwidth]{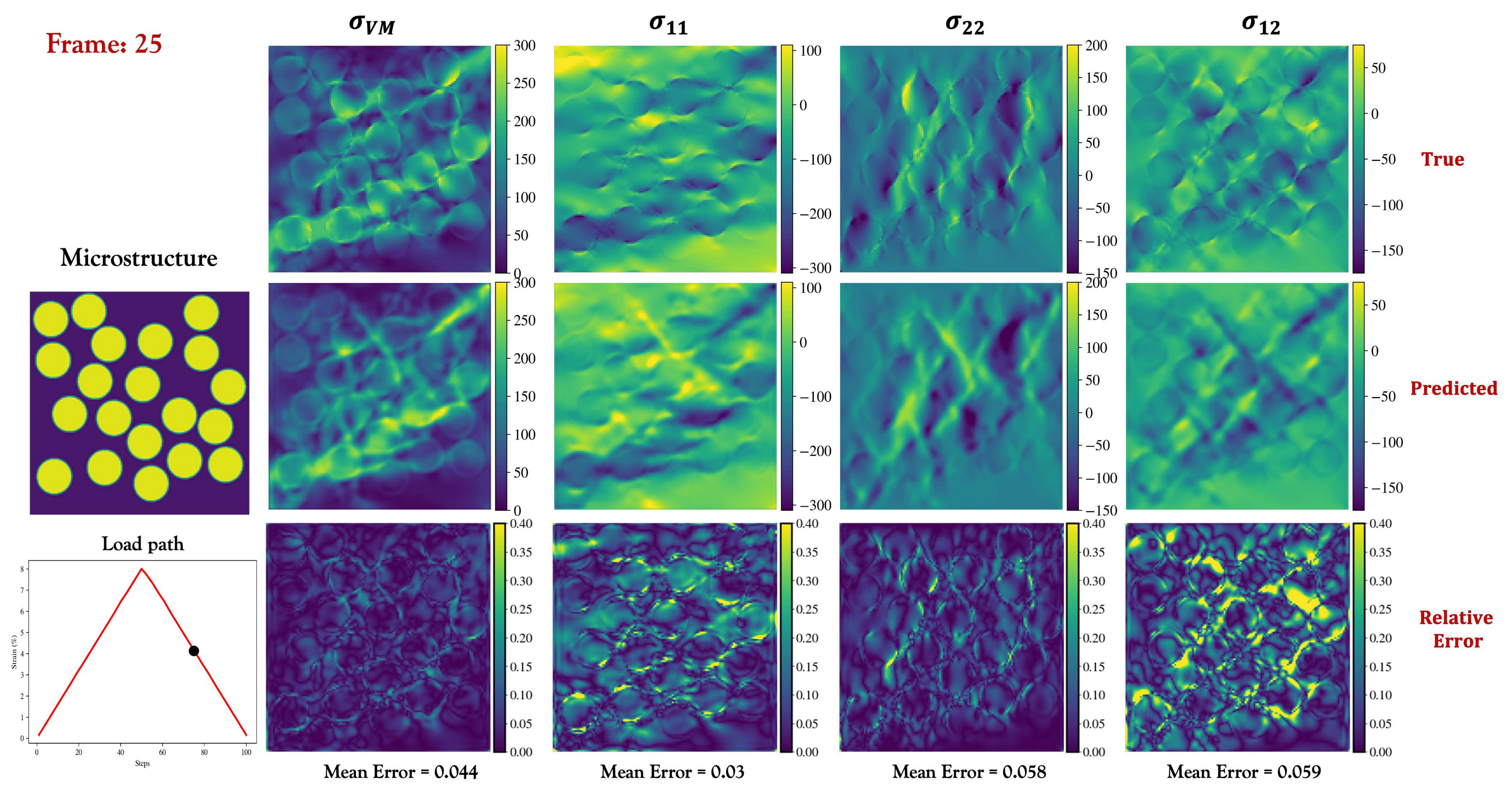}
         \caption{Comparison of the predicted and the true stress maps in the unloading stage (Frame: 25)}
         \label{fig:stress25}
          \hfill
\end{figure}

 U-Net predictions of the stress maps at frame 10 (loading) are shown in Fig.~\ref{fig:stress10}. These results show good agreement between the predictions and the ground truth FE analyses, with local RE ranging from 0 to 40\% and a mean RE less than 3.3\% for all stresses. This error is a little higher than that observed for the strains, but still remains quite small. 
 There is more visible error in the unloading stage at frame 25 (Fig.~\ref{fig:stress25}). Again, the RE ranges from 0 to 40\% but the higher error regions are more widely dispersed, as indicated by a mean RE as high as 5.9\% in the case of the shear stress ($\sigma_{12}$). 
 In particular, the error plots show that some of the localized regions around the fiber-matrix interfaces exhibit higher errors. The high error around the fiber-matrix interface is due in part to the discretization of the circular fibers with a coarse rectangular grid in order to generate the images. Such interfaces present a challenge, even with standard FE simulations.  
 The unloading stage presents a particularly difficult challenge, as the average stress, and therefore the range of stresses, is decreasing. The ML model is trained for all stresses at all load steps simultaneously, so that capturing the relatively small variations in stresses at the late unloading steps is difficult. This effect is particularly noticeable in the prediction of $\sigma_{12}$, 
 which has a smaller range of stresses than the other stress components throughout the load path. 


\begin{figure}[t]
         \includegraphics[width=1\textwidth]{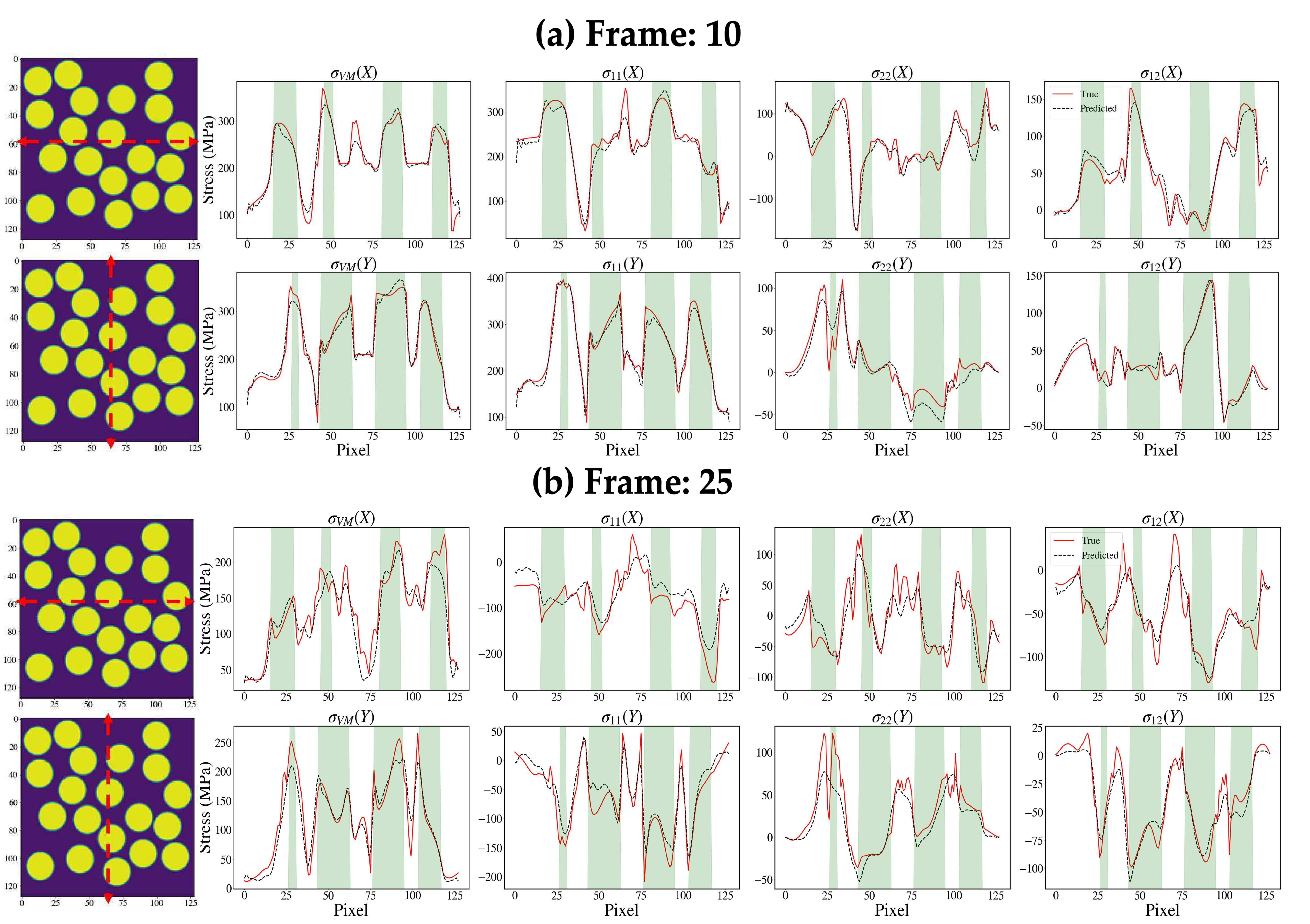}
         \caption{Variation of stresses along the midsection. (a) Frame 10 (b) Frame 25 (The green zones signify the regions where the line intersects the fiber)
         }
         \label{fig:stress_slice}
          \hfill
\end{figure}

Some of these errors are easier to visualize by considering the stresses along the two perpendicular lines for frames 10 and 25. The predicted stresses and the true stresses on the loading path (frame 10) agree with each other in most cases (Fig.~\ref{fig:stress_slice}a). 
The most significant errors are found in the regions near fiber boundaries. The predicted stresses exhibit more error on the unloading path (Fig.~\ref{fig:stress_slice}b), with error found not only near fiber-matrix interfaces but also within fiber and/or matrix interior regions. 

\begin{figure}[h!]
\includegraphics[width=1\textwidth]{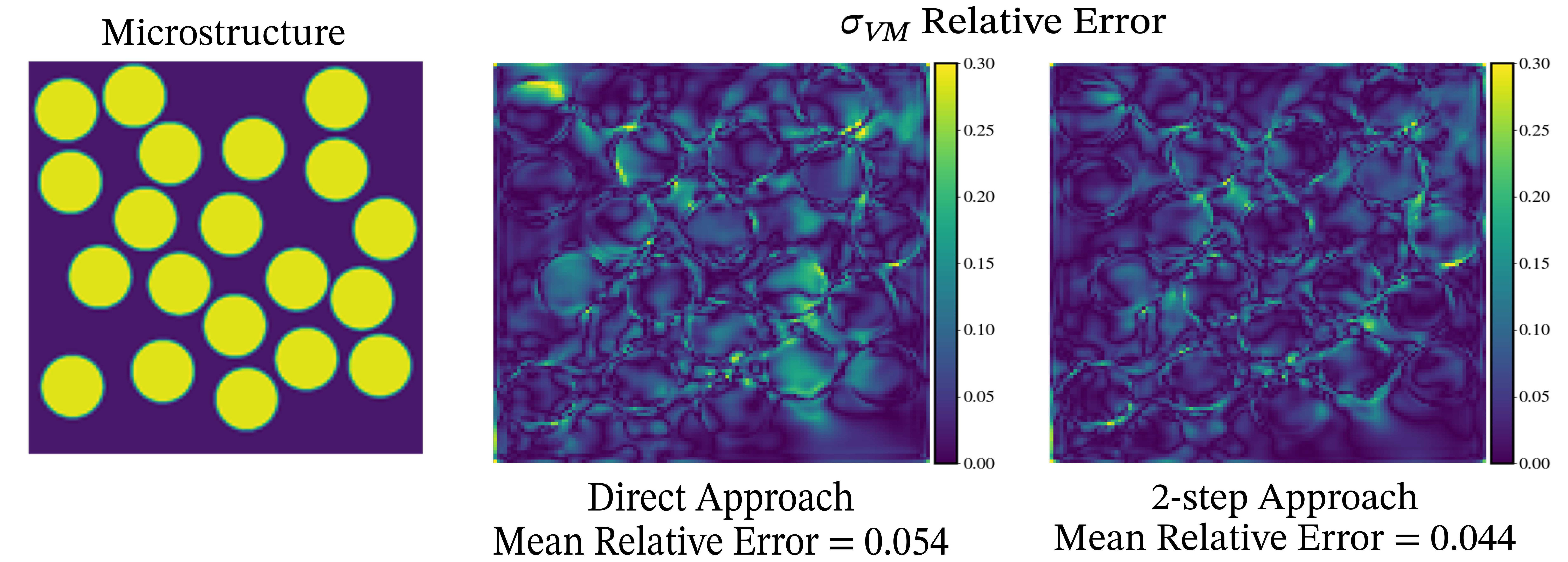}
         \caption{Comparison of the direct approach and the 2-step approach of von-Mises stress prediction at frame 25 on the unloading path. 
         }
         \label{fig:box-comparison-stress}
          \hfill
\end{figure}

As mentioned earlier, it is possible to predict the stresses based directly on the microstructure and the load factors as input, using the U-Nets in a similar way as for the strains. Another approach is to use a 2-step approach, in which the U-Net-predicted strains are an input to the stress U-Net, along with the microstructure. In Fig.~\ref{fig:box-comparison-stress}a the RE maps for the von-Mises stress (on the unloading path) are given. The error map for the direct approach shows that there are more regions with a higher error than the 2-step approach. 
The median error for frame 25 shows that the direct approach has a higher mean RE ($5.4\%$) than the 2-step approach ($4.4\%$). 
Based on these observations, the 2-step approach is used in all other analyses in this paper. 

\subsection{Sample to sample variability in mean relative error}
\begin{figure}[!h]
  \centering
         \includegraphics[width=0.9\textwidth]{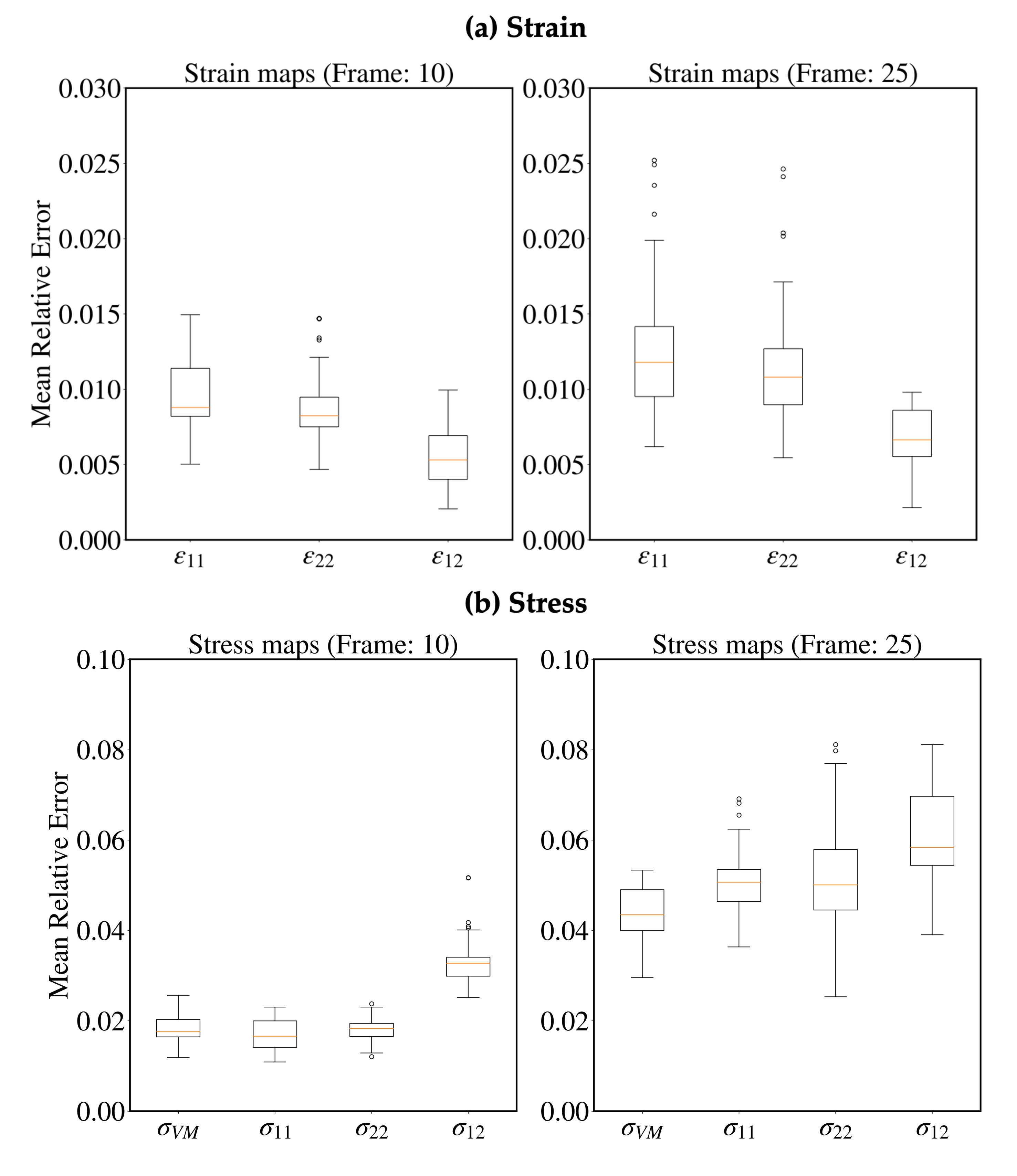}
         \caption{Box-plots for the mean relative error in the test dataset for (a) Strains (b) Stresses 
         }
         \label{fig:box_plot}
          \hfill
\end{figure}

The RE, as defined in Eq.~\ref{Error}, provides a measure of the normalized pixel-by-pixel error that can be averaged over the image to yield a mean RE. This mean RE varies from one instantiation of microstructure to the next, which is not reflected in the values reported in Figs.~\ref{fig:strain10},~\ref{fig:strain25},~\ref{fig:stress10}, or \ref{fig:stress25}. Using all of the test data from this analysis, the resulting ensembles of mean RE are depicted in Fig.\ref{fig:box_plot}a and Fig.\ref{fig:box_plot}b. 
The box plots for strain maps in Fig.\ref{fig:box_plot}a show that the median values for the mean errors are quite low, less than $2\%$ even in the worst case at Frame 25. The sample-to-sample variations in this mean RE are also quite small. The box plots for stress maps show an increase in the mean RE, with an increase in both the median and the scatter when going from the loading to the unloading path. 
On the loading path (Frame 10) the median value of the mean RE remains lower than $4\%$. However, the unloading path (Frame 25) shows a median mean RE of around $6\%$ for $\sigma_{12}$. 
An intriguing observation from this plot is that although $\sigma_{11}$, $\sigma_{22}$, and $\sigma_{12}$ exhibit a significant increase in mean RE when going from loading to unloading, the RE in the von-Mises stress $\sigma_{VM}$ remains relatively low. Von-Mises stress is a function of the other stress components - $\sigma_{11}$, $\sigma_{22}$, and $\sigma_{12}$ - suggesting there may be some sort of error balancing occurring in the von-Mises formulation of stress.

\subsection{Quantification of average error for entire load path}
While it is hard to visualize the full stress and strain fields at all 32 load frames, it is useful to address the evolution of average error in these quantities over the entire load path. This is achieved using two metrics:  $R^2$ values, and average stress-strain curves.

\subsubsection{Evaluation of the $R^2$ values for strain and stress maps for each frame}
In this section, the fit of the predicted values $z_{pred}^{(mn)}$ and the true values $z_{true}^{(mn)}$  with the $y=x$ straight line is determined using the coefficient of determination ($R^2$). $R^2$ can be calculated as,
\begin{equation}
    R^2=1-\frac{\Delta z}{Z}
        \label{eq:2}
\end{equation}
where,
\begin{equation}
    \Delta z=\mathlarger{\mathlarger{\sum}}_{m,n}^{N} (z_{true}^{(mn)}-z_{pred}^{(mn)})^2
\end{equation}
\begin{equation}
    Z=\mathlarger{\mathlarger{\sum}}_{m,n}^{N} (z_{true}^{(mn)}-\bar{z}_{true})^2
\end{equation}
and
\begin{equation}
    \bar{z}_{true}=\frac{1}{N^2}\mathlarger{\mathlarger{\sum}}_{m,n}^{N} z_{true}^{(mn)}    
\end{equation}
\begin{figure}[!h]
  \centering
         \includegraphics[width=1\textwidth]{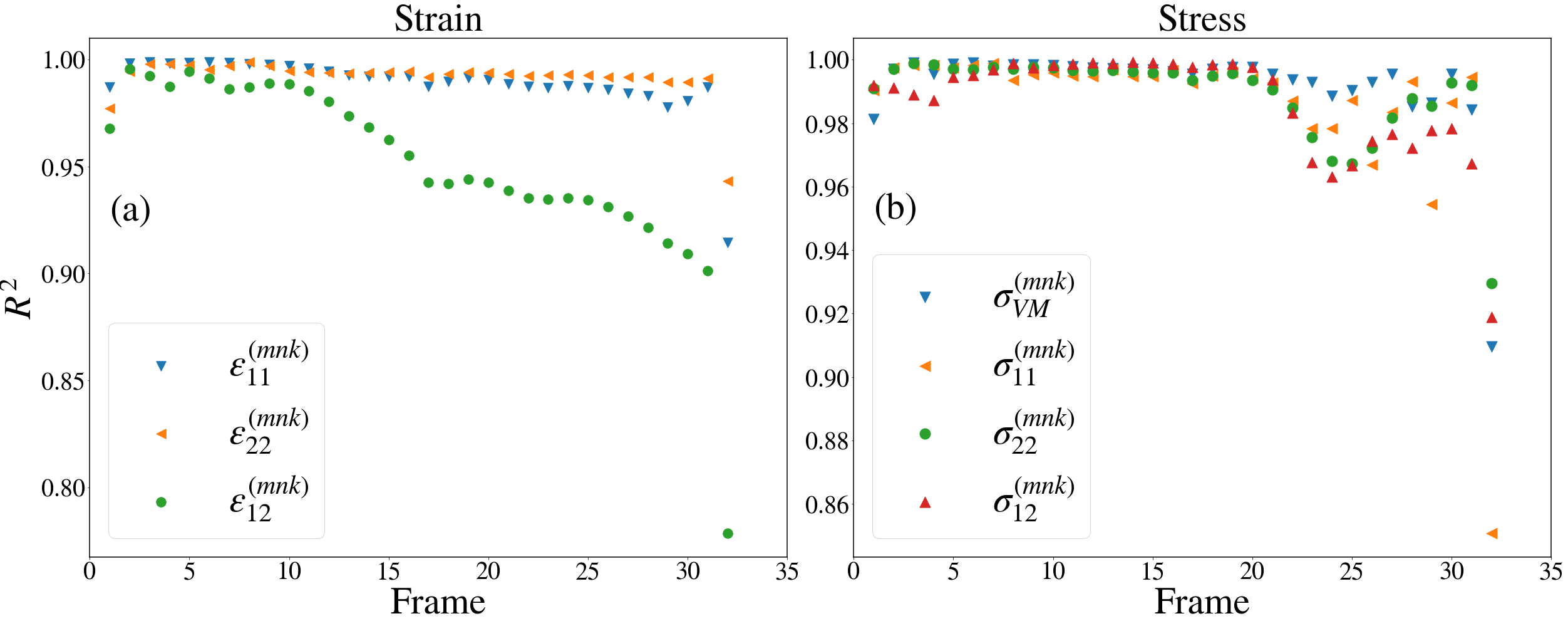}
         \caption{Variation of the $R^2$ values with the image frames. Frames refer to each image along the third dimension and each frame is related to the sampled time steps from the load path in Fig.~\ref{fig:Actual load path} (a) Strain Maps (b) Stress Maps}. 
         \label{fig:r2_score}
          \hfill
\end{figure}
Therefore, $R^2$ is related to the spread of the predicted pixel values from the true values. Fig.~\ref{fig:r2_score}a shows $R^2$ for the strain predictions, as a function of load frame. $R^2$ remains almost constant for both axial strains, with the exception of the final frame, at which there is a sharp drop. For the shear strain $\epsilon_{12}$, there is a steady decrease in $R^2$ with load frame, but all the values remain above 0.90 except for the last frame which has a value of 0.73. 
Fig.~\ref{fig:r2_score}b shows $R^2$ for the stress predictions, which exhibits a similar behavior as the strain predictions. For most of the frames, this value is relatively constant and remains close to 1. After unloading at around Frame $20$, the $R^2$ values begin to decrease, in particular for $\sigma_{11}$ and $\sigma_{22}$. Even in the worst case of $\sigma_{11}$ at the final load frame, however, the $R^2$ value is around 0.85. 

\subsection{Average stress-strain response of the microstructure}
\begin{figure}[t]
  \centering
         \includegraphics[width=1\textwidth]{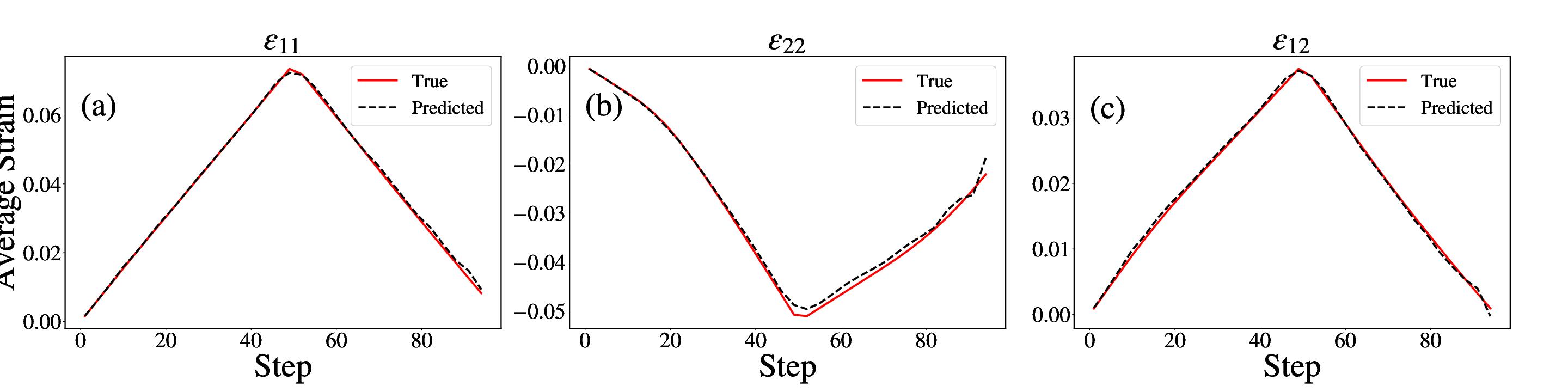}
         \caption{Variation of average strain with time step}
         \label{fig:avg_strain}
          \hfill
\end{figure}
\begin{figure}[!htb]
  \centering
         \includegraphics[width=\textwidth]{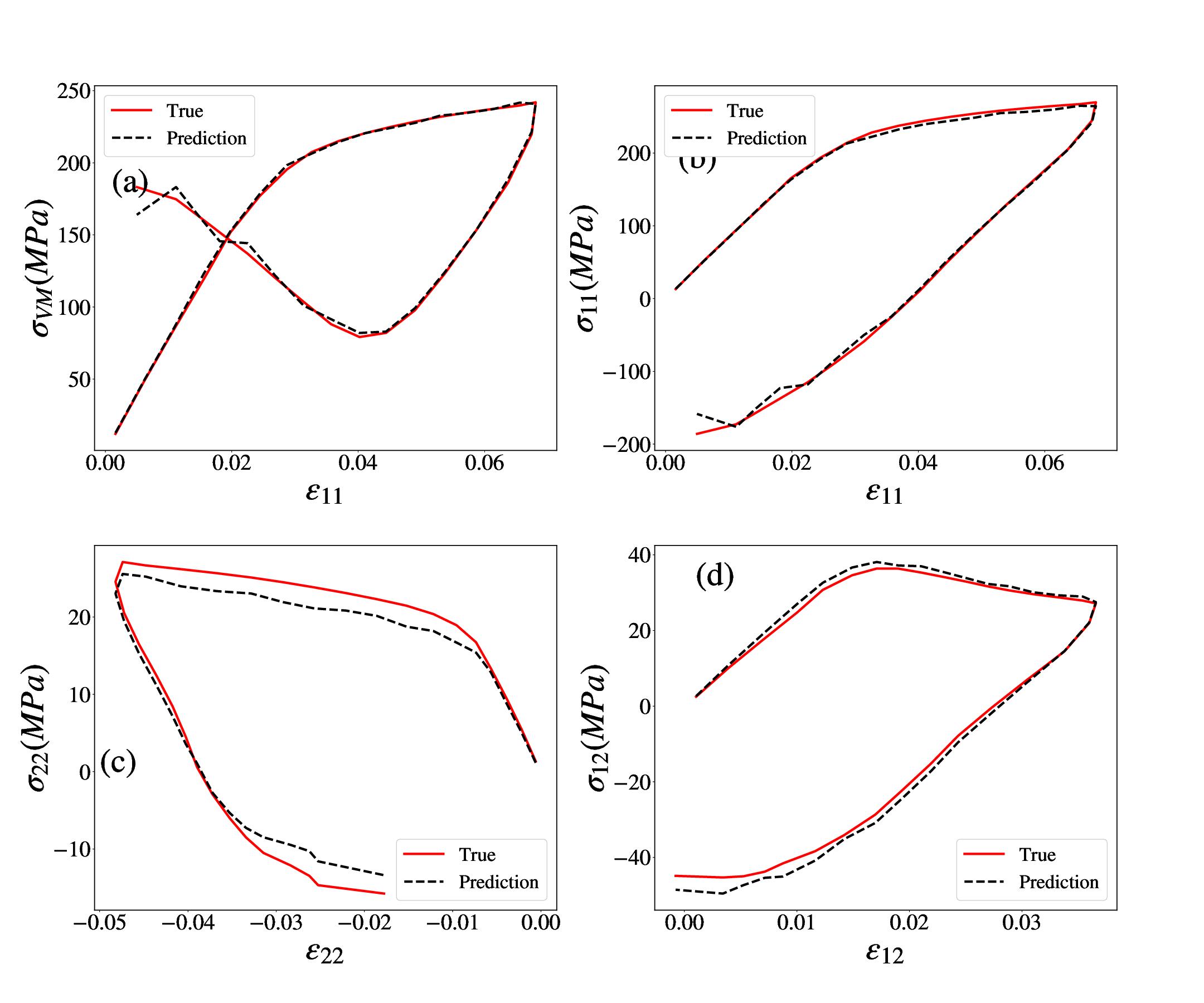}
         \caption{Average stress-strain relationship of the microstructure}
         \label{fig:avg_stress_strain}
          \hfill
\end{figure}
Given that these ML models might be expected to predict average mechanical behavior, it is useful to consider the average strains and stresses predicted by the model. Fig.~\ref{fig:avg_strain} shows the variation of the average strains in the microstructure over the total duration of loading and unloading, comparing the predicted and true values. There is excellent agreement between the average strains from the predicted maps (from UNet\_EIJ) and the average strains from the true maps from FE analysis.  Fig.~\ref{fig:avg_stress_strain} shows the average stress-strain behavior throughout the load path. Considering the von-Mises stress, yielding is observed at the matrix yield stress of $210 MPa$, with subsequent hardening behavior due to the presence of the elastic fibres. This expected response is captured accurately by the U-Net predictions. In fact, there is excellent agreement in all of these curves, with the possible exception of $\sigma_{22}$, which shows some discrepancies, likely because this stress is of smaller magnitude relative to the other stresses and therefore more difficult to capture. 

\section{Conclusions}\label{sec:conclusions}
In this study, we aimed to develop a data-driven method to predict the stresses and strains in a two-phase elastoplastic material using 3D convolutions in a U-Net type framework. The approach is orders of magnitude faster than the equivalent FE analysis, which is key to multiscale modeling and/or stochastic simulation of local material response. 3D convolutions are capable of addressing the temporal correlations in a series of stress or strain maps at different instants of the load path, which are akin to a sequence of images. Based on these analyses, we can draw the following conclusions:
\begin{enumerate}
  \item 3D U-Net can predict stresses and strains in an inelastic medium with arbitrary fiber placement, under a given load path. It can therefore produce  accurate results in the plastic domain based on limited prior information.
  \item There is excellent agreement between U-Net and FEM in predicting local strains. The maximum mean RE observed for all strain maps was approximately $1.1\%$ on the loading path and $1.6\%$ on the unloading path. 
  \item There is also very good agreement between U-Net and FEM in predicting local stresses. The maximum mean RE observed for all stress maps was approximately $3.3\%$ on the loading path and $5.9\%$ on the unloading path. 
  \item Upon calculating the mean RE for all the unseen microstructures, it was observed that the strain predictions are highly accurate, with the maximum median error being less than $1.5\%$ in loading as well as unloading cases. The stresses show somewhat of an increase in the median $RE$ in the unloading cases, with a maximum around $6\%$.  
  \item The predicted average stress-strain response shows excellent agreement with the FE prediction of the stress-strain response.
  \item $R^2$ values are calculated for strains in the range of $0.73 - 0.98$ and for stresses in the range of $0.85 - 0.98$, demonstrating excellent agreement between the U-Net and FE predictions.
\end{enumerate}
\indent Although there are some localized regions of high error in the predicted stresses, the model captures local changes in inelastic stresses and strains with excellent accuracy, at a small fraction of the computational cost required by FE analysis. This work did not consider changes in fiber sizes, shapes and/or volume fraction, which is left as a future direction. Prior work in this regard on elastic stresses suggests that the U-Net model might be capable of representing these changes to the microstructure. A more challenging future direction is to consider other types of microstructures, such as polycrystalline materials. Given the larger number of possible characteristics represented at each pixel in the microstructure (e.g., different grain orientations), as opposed to the two possibilities in a fiber-reinforced composite (i.e., fiber or matrix), more data might be required to train the model effectively.

Another consideration is the trade-off between training time and analysis time. Once trained, the U-Net model provides a very rapid and accurate assessment of local stresses and strains; however, the training process itself takes some computational effort and also requires data from the finite element analyses. The choice on whether to implement such an ML model depends on the number of model evaluations one is expecting to perform. For example, in a multiscale model that has a unique microstructure underlying each integration point of a larger scale model, there are expected to be a large number of micro-scale model evaluations. In such a case, the U-Net model is ideal. On the other hand, if one only wants a small number of micro-scale evaluations, then the data collection and training process may be more cumbersome than the time associated with simply running FE analyses of each individual case. Formulation of specific quantitative metrics to identify this trade-off point will be highly problem dependent, and therefore no general rule of thumb is available here.

Finally, it is important to note that the current work is an extremely powerful tool for predicting local stresses and strains in a microstructure, under a fully known load path. The enormous increase in computational speed provided by the U-Net promises to enable efficient multi-scaling and/or uncertainty quantification for complex mechanics models. However, it is important to recognize that in a multiscale mechanics context, these individual microstructures might be part of a larger scale model in which the applied displacements/strains on each microstructure are associated with individual nodes or integration points. In this case, the load path applied to each microstructure could vary from one point to the next. Therefore, a clear next step for such U-Net models is to generalize the model for multiple load paths. 

\section{Acknowledgements}
Research was sponsored by the Army Research Laboratory and was accomplished under Cooperative Agreement Number W911NF-22-2-0014 (Artificial Intelligence for Materials Design) and Cooperative Agreement Number W911NF-23-2-0062 (AI-Driven Integrated and Automated Materials Design for Extreme Environments). The views and conclusions contained in this document are those of the authors and should not be interpreted as representing the official policies, either expressed or implied, of the Army Research Office or the U.S. Government. The U.S. Government is authorized to reproduce and distribute reprints for Government purposes notwithstanding any copyright notation herein.
\bibliography{mybibfile}

\end{document}